\documentclass[%
 aip,
 amsmath,amssymb,
 reprint,%
]{revtex4-1}

\usepackage{graphicx}
\usepackage{dcolumn}
\usepackage{bm}

\usepackage[utf8]{inputenc}
\usepackage[T1]{fontenc}
\usepackage{mathptmx}
\usepackage{subcaption}
\usepackage{xcolor}
\begin{document}

\preprint{AIP/123-QED}

\title[Sample title]{Attenuating surface gravity waves with mechanical metamaterials}

\author{F. De Vita}
 \email{francesco.devita@poliba.it}
 \affiliation{Dipartimento di Meccanica, Matematica e Management, Politecnico di Bari, Via Re David 200 – 70125 Bari, Italy}

\author{F. De Lillo}
\affiliation{Dipartimento di Fisica, Universit\`a degli Studi di Torino, Via Pietro Giuria 1, 10125 Torino, Italy}

\author{F. Bosia}
\affiliation{DISAT, Politecnico di Torino, Corso Duca degli Abruzzi 24, 10129 Torino, Italy}

\author{M. Onorato}
\affiliation{Dipartimento di Fisica, Universit\`a degli Studi di Torino, Via Pietro Giuria 1, 10125 Torino, Italy}

\date{\today}

\begin{abstract}
  Metamaterials and photonic/phononic crystals have been successfully developed in recent years to achieve advanced 
  wave manipulation and control, both in electromagnetism and mechanics. However, the underlying concepts are yet to 
  be fully applied to the field of fluid dynamics and water waves. Here, we present an example of the interaction 
  of surface gravity waves with a mechanical metamaterial, i.e. periodic underwater oscillating resonators. In particular, 
  we study a device composed by an array of periodic submerged harmonic oscillators whose objective is to absorb wave 
  energy and dissipate it inside the fluid in the form of heat. The study is performed using a state of the art direct 
  numerical simulation of the Navier-Stokes equation in its two-dimensional form with free boundary and moving bodies. 
  We use a Volume of Fluid interface technique for tracking the surface and an Immersed Body method  for the fluid-structure 
  interaction. We first study the interaction of a monochromatic wave with a single oscillator and then add up to four 
  resonators coupled only fluid-mechanically. We study the efficiency of the device in terms of the total energy dissipation 
  and find that by adding resonators, the dissipation increases in a non trivial way. As expected, a large energy attenuation 
  is achieved  when the wave and resonators are characterised by similar frequencies. As the number of resonators is increased, 
  the range of attenuated frequencies also increases. The concept and results presented herein are of relevance for applications 
  coastal protection.
\end{abstract}

\maketitle

\section{Introduction\label{sec:introduction}}

In recent years, the field of mechanical metamaterials and phononic crystals has seen a rapid development and 
captured increasing interest  \cite{deymier2013introduction}. They are engineered materials that have been 
developed to alter the standard properties of wave propagation such as dispersion, 
refraction or diffraction. Metamaterials are usually arranged in periodic patterns, at 
scales that are comparable or smaller than the wavelengths of the phenomena they influence. The simplest 
effect is that when waves propagate in a periodic structure the dispersion relation displays 
banded structures with frequency regions that are forbidden, called band gaps. This effect, 
for example, can be obtained in phononic biatomic materials \cite{hussein2014dynamics}.
The concept of metamaterials was first developed in the field of optics \cite{pendry2001} and 
later extended to phononic crystals and elastic waves \cite{hussein2014dynamics}.
Some work on the interaction of gravity waves with a 
macroscopic periodic structure (a sinusoidal floor) was already considered 
(see \cite{davies1984surface,hara1987bragg}). Results indicated the existence of a mechanism
of resonant Bragg reflection occurring when the wavelength of the bottom undulation is one 
half the wavelength of the surface wave. Further studies on the interactions of waves with 
periodic structures can be found in \cite{hu2003complete,hu2011negative,kar2020bragg}. 
Other examples of wave manipulation properties, for example cloaking, can be obtained by employing 
an engineered elastic buoyant carpet placed on water \cite{zareei2016cloaking} or by a radial
arrangment of vertical cylinders \cite{zhang2020broadband}.

The interaction of ocean waves with structures is a long standing problem in fluid mechanics 
\cite{mei1989applied}. A theoretical approach based on the direct use of the equations of 
motion, even in their simplified version, is not always feasible, especially when geometries 
are not simple and bodies are moving because of hydrodynamical forces. 
In the latter cases, an experimental approach is not easy as the measurement of pressures and 
of the velocity field around the moving bodies may not be straightforward. Numerical methods, 
despite their complexity, often offer an important alternative for studying 
wave-structure interaction and  designing structures. With respect to standard fluid 
mechanics, the main complication arises because of the presence of a free surface which 
substantially increases the difficulty of the numerical treatment. Some studies in the 
literature assume that the flow is irrotational and inviscid so that the potential flow 
equations can be solved and forces are limited to pressure \cite{heikkinen2013,abbasnia2018}. 
However, when the goal is to study the overall effect of wave attenuation and the energy 
dissipated in the bulk of the fluid, vorticity and viscosity cannot be neglected and the full 
Navier-Stokes equations need to be computed: recent works 
\cite{jin2018,xu2019performance} have provided evidence that viscosity plays an important role, 
especially close to resonant conditions. To this end, Direct 
Numerical Simulation (DNS) of a free surface flow interacting with a structure represents 
a powerful tool that can provide a detailed representation of the flow field and of the 
fluid-structure interaction.

In this work, we consider the interaction of gravity waves with a  periodic structure 
composed by ``internal'' resonators, i.e. waves interact with submerged 
harmonic oscillators which are coupled only fluid-mechanically. 
To begin with, we work in a two dimensional framework; therefore, strictly speaking, our waves are 
characterised by infinitely long crests and the oscillators are cylinders whose
axes are parallel to the crests. We solve the full Navier-Stokes system of
equations coupled with the Volume of Fluid (VoF) method for the interface
tracking and the Immersed Boundary Method (IBM) for the fluid-structure
interaction. The cylinders undergo the hydrodynamic forces (pressure and
viscous stress) of the wave motion and an elastic force which tends to restore
the system back to the equilibrium position. The analysis has been conducted
for a variable number of resonators per wavelength and varying their natural
frequency. It is worth mentioning that the
system we are considering is similar to systems for wave energy conversion, on
which there is a rich literature ranging from point-absorbers
\cite{li2012synthesis,zurkinden2014non,xu2019performance} to a full modeling of
the solid structure \cite{heikkinen2013,anbarsooz2014}. However, the focus here
is on the interaction of a wave with a periodic structure rather than the
conversion of energy from a single oscillator.

\section{Methodology\label{sec:methodology}}
\subsection{The numerical method}
We solve the full Navier-Stokes system of equations
\begin{gather}
    \rho(\partial_t \mathbf{u} + \mathbf{u}\cdot\nabla \mathbf{u}) = -\nabla p + 
    \nabla \cdot (\mu \mathbf{D}) + \rho\mathbf{g} + \mathbf{f} \\
    \nabla \cdot \mathbf{u} = 0
\end{gather}
with $\mathbf{u} = (u,w)$ the velocity field, $p$ the pressure field, $\mathbf{D}$ the
deformation tensor defined as $D_{ij} = (\partial_i u_j + \partial_j u_i)/2$, $\mathbf{g}$
the gravity vector and $\mathbf{f}$ the IBM force which enforces the no-slip boundary
condition at the solid boundary. The material properties $\rho$ and $\mu$ are related to the
volume fraction field $\mathcal{F}(\mathbf{x},t)$ as 
\begin{gather}
    \rho({\mathcal{F}}) = \mathcal{F}\rho_1 + (1-\mathcal{F})\rho_2 \\
    \mu({\mathcal{F}}) = \mathcal{F}\mu_1 + (1-\mathcal{F})\mu_2,
\end{gather}
where $\rho_1$, $\rho_2$, $\mu_1$ and $\mu_2$ are the density and viscosity of the two fluids;
the volume fraction field (defined as the volumetric ratio of the two fluids in each computational cell)
is advected by the flow with the following equation
\begin{equation}
    \partial_t \mathcal{F} + \nabla \cdot (\mathcal{F}\mathbf{u}) = 0.
\end{equation}
The motion of the resonators is given by Newton's law
\begin{equation}
    \label{eqn:Newton}
    m_i \frac{d^2X_i}{dt^2} + \kappa_i (X_i - X_{0,i})= F_i
\end{equation} 
where $X_i$ is the position of the centre of mass of the \emph{i-th} resonator, $m_i$ its mass,
$\kappa_i$ is the elastic constant, $X_{0,i}$ the equilibrium position and $F_i$ the integral
of the hydrodynamic forces acting on it.This force is computed by integrating 
the pressure ($p$) and the viscous stress tensor ($\boldsymbol{\tau}$) over the surface of the solid body as follows:
\begin{equation}
  \label{eqn:force}
\mathbf{F} = \int_S\left(\boldsymbol{\tau} - p\mathbf{I}\right) \cdot \mathbf{n} dS.
\end{equation}
By computing the force in this way, all terms 
typically used in the description of point-absorbers (such as viscous damping and radiation damping)
are included, and eq.\eqref{eqn:force} provides a more general and accurate description of the solid body motion.

Modelled in this way, the resonator has a natural
frequency $\omega_r = \sqrt{k/m}$. In the real system, resonators would correspond to reversed pendula
anchored at the bottom; for waves of small amplitude, as in this study, the vertical motion of the resonators 
can be neglected, hence, we solve equation \eqref{eqn:Newton} only for the horizontal motion, with 
$F_i$ being the horizontal component of the integral of the hydrodynamic loads acting on the \emph{i-th} resonator;
the motion in the vertical direction is set to zero.

\begin{figure}
  \centering
  \includegraphics[width=0.8\columnwidth]{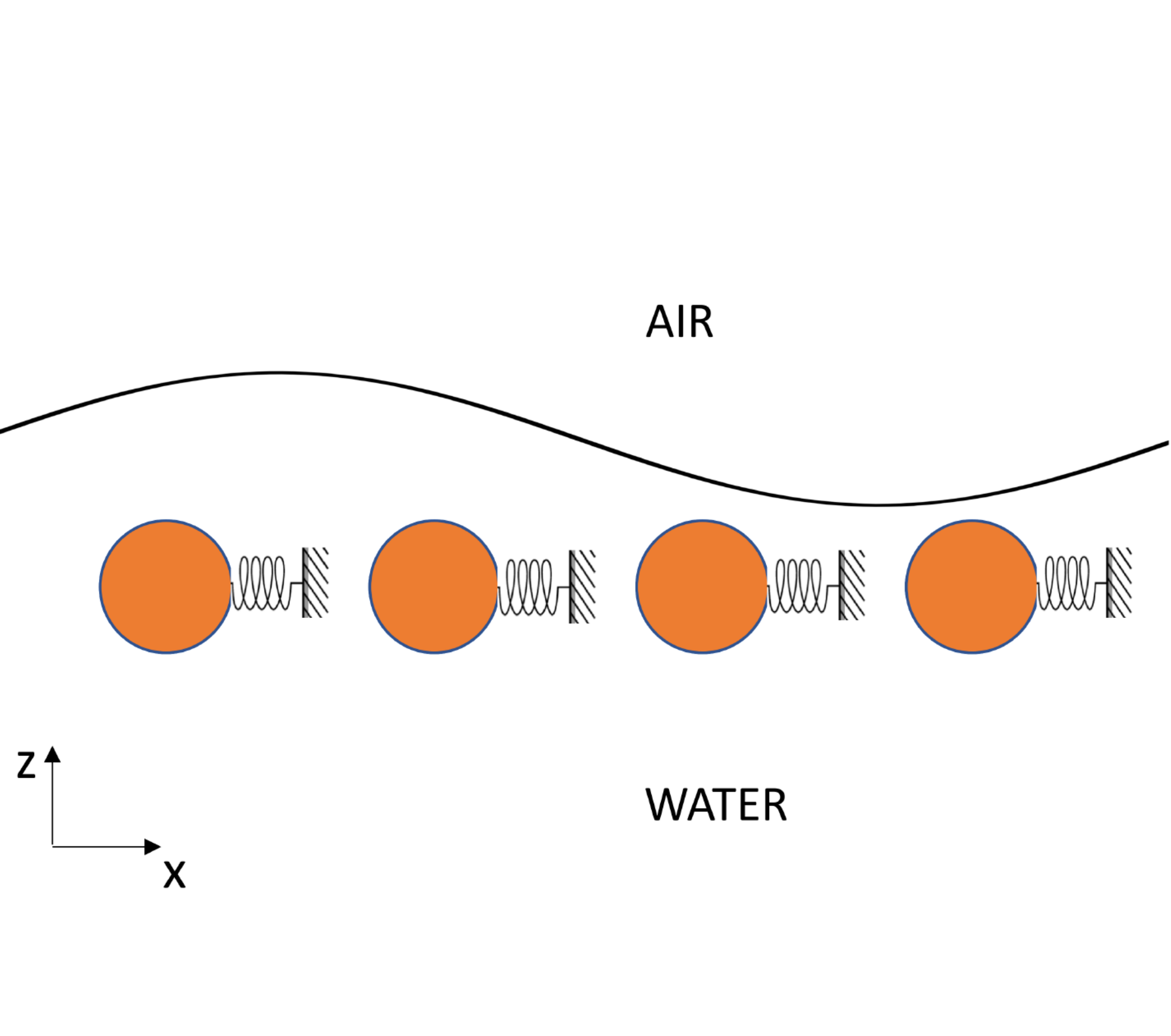}
  \caption{Sketch of the periodic structure of resonators interacting with a surface gravity
     wave\label{fig:disegno}}
\end{figure}

The Navier-Stokes equations are advanced in time using a 2-{\it nd} order Adams-Bashforth scheme and  a fractional step 
method is employed \cite{Kim1985} for the 
coupling with the pressure; the resulting Poisson
equation for the pressure is solved employing a Fast Direct Solver. All derivatives are discretized 
with a second order central difference scheme apart for the diffusion term in the Navier-Stokes 
equations, for which a WENO scheme is used \cite{Shu2009}. The IBM is implemented 
using the direct forcing approach \cite{Fadlun2000} with interpolations performed in the direction 
normal to the interface. For the fluid-structure interaction, a strong coupling is adopted with an 
iterative solver based on the Hamming method \cite{Hamming1959}. The solver is limited 
to non-deformable solid bodies, which allows for more efficient computations. A detailed description of the solver
with validations and preliminary results can be found in \cite{JCPIBMVOF}. 
A sketch of the periodic structure of four resonators immersed in a fluid and forced by surface gravity waves is 
displayed in  Fig. \ref{fig:disegno}.

\begin{figure*}[th]
  \centering
  \begin{subfigure}{0.49\textwidth}
    \includegraphics[width=\textwidth]{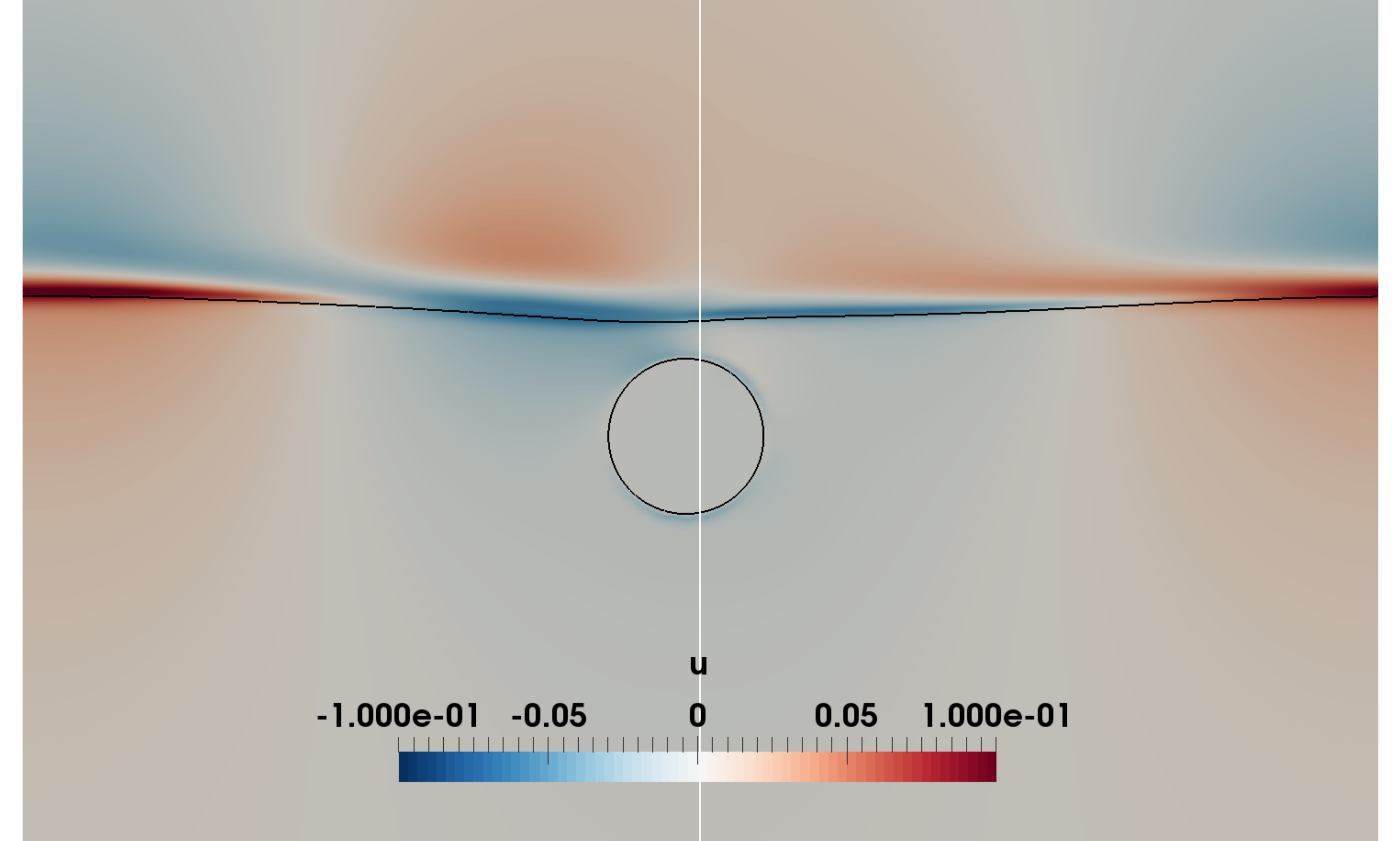}
    \caption{$t/T = 0.5$.}
  \end{subfigure}
  \begin{subfigure}{0.49\textwidth}
    \includegraphics[width=\textwidth]{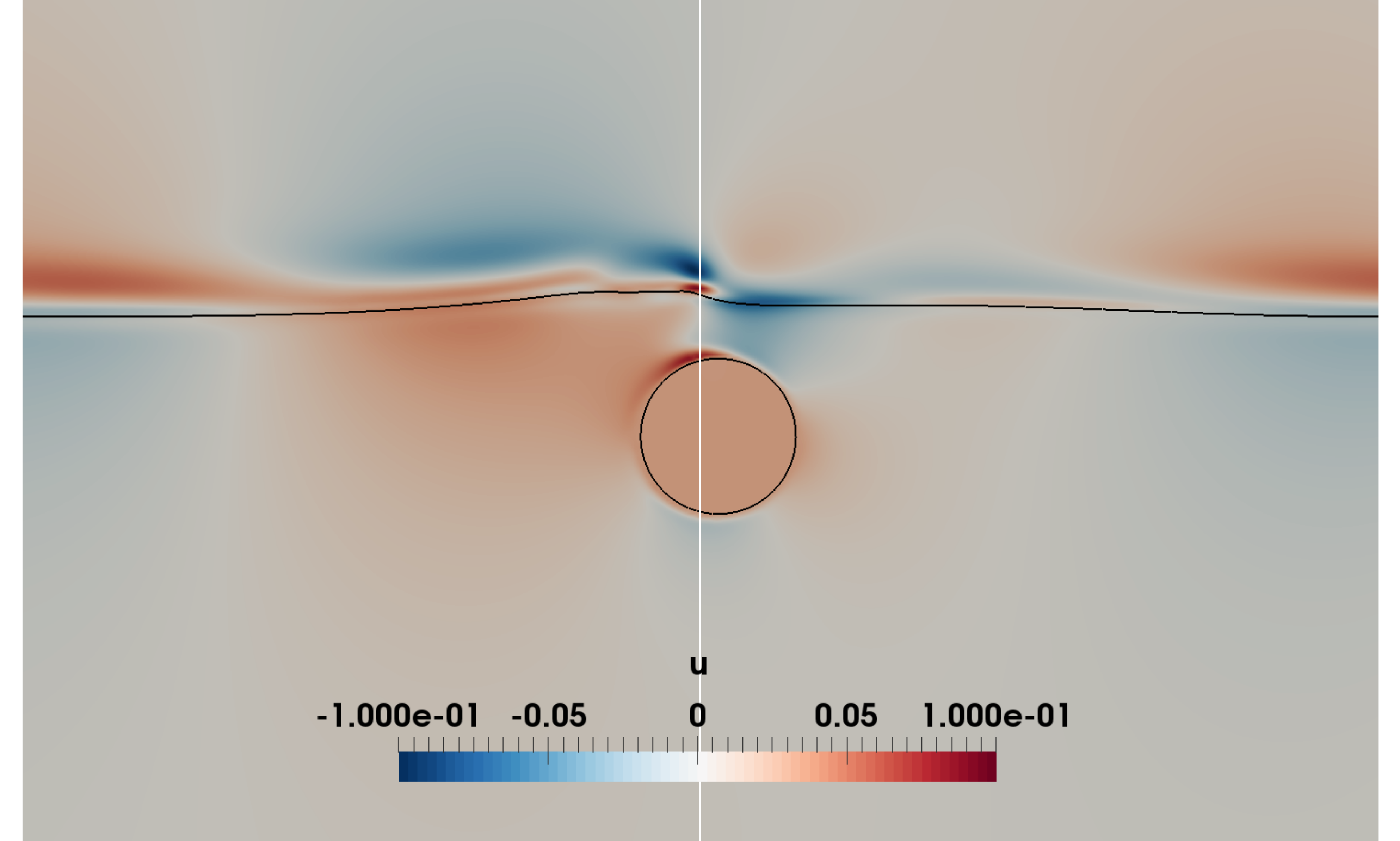}
    \caption{$t/T = 1$.}
  \end{subfigure}
  \begin{subfigure}{0.49\textwidth}
    \includegraphics[width=\textwidth]{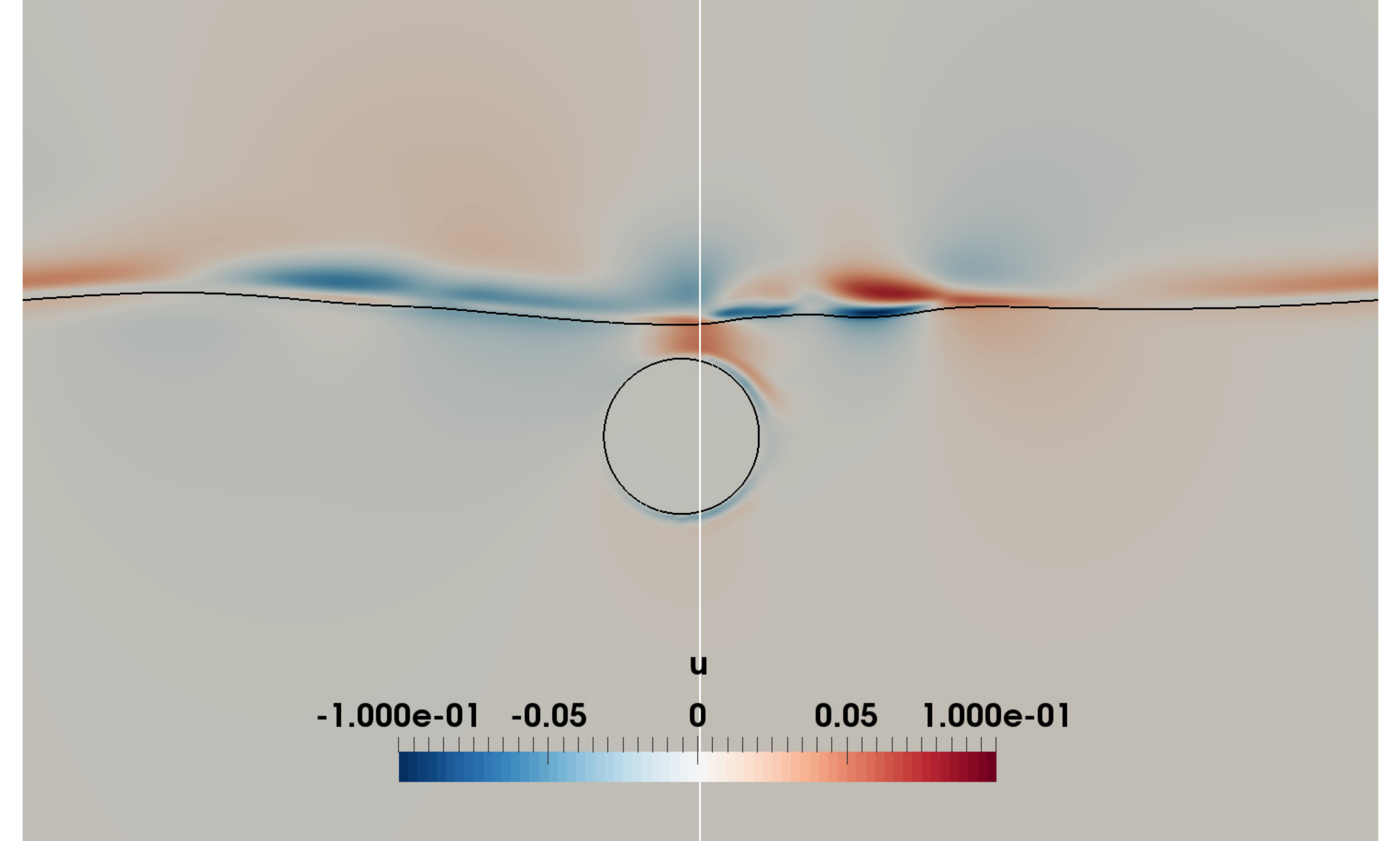}
    \caption{$t/T = 1.5$.}
  \end{subfigure}
  \begin{subfigure}{0.49\textwidth}
    \includegraphics[width=\textwidth]{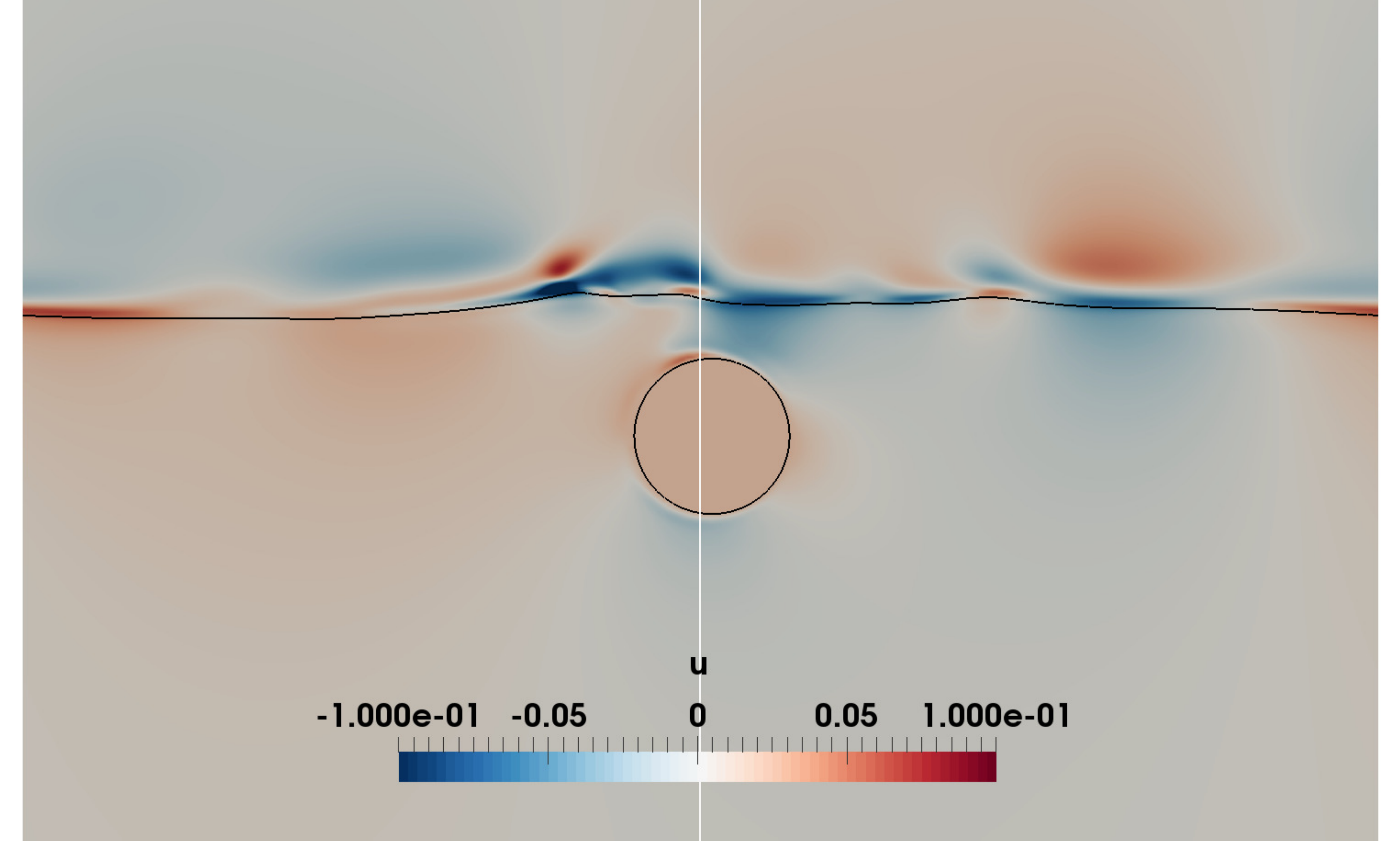}
    \caption{$t/T = 2$.}
  \end{subfigure}
  \caption{Snapshot of the horizontal component $u$ of the velocity field at four instants  of time for the case 
  with $\Omega = 1$. The vertical white line is located at the center of the domain and it is reported as 
  reference to underline the motion of the resonator. (Multimedia view)\label{fig:1pvis}}
\end{figure*}

\subsection{Initial conditions}

The initial wave profile $\eta$ and velocity field $\mathbf{u} = (u,w)$ are taken from linear theory
and are, respectively:
\begin{equation}
    \eta(x,0) = a \cos(kx)
\end{equation}
and
\begin{gather}
    u(x,0) = a \omega e^{kz} \cos(kx) \\
    w(x,0) = a \omega e^{kz} \sin(kx),
\end{gather}
with $a$ the wave amplitude, $k = 2\pi/\lambda$ the wavenumber, $\lambda$ the wavelength
and $z = 0$ the still water level, and $z$ pointing upward. The wave frequency $\omega$ is
given by the dispersion relation for water waves in deep water $\omega = \sqrt{gk}$. The
initial velocity field in air is equal to that in water with the horizontal component $u$ with
a negative sign. To avoid high shear stress across the interface at the beginning of the
simulation, the initial volume fraction is filtered with a bilinear interpolation which results
in a spread of the interface over three cells. Note that this operation is performed only for
the initial profile. The computational domain is a square of lateral size $\lambda$, the radius of the
resonators is $r = 0.057 \lambda$ and the distance from the centre
of mass of the resonators and the still water level is $d = 0.094 \lambda$.  The Reynolds number based on the
wavelength and phase speed is set to $Re = \rho g^{1/2}\lambda^{3/2}/\mu = 10^5$, with 
$\rho$ and $\mu$ the density and viscosity of the high density phase. All simulations are performed with a grid of 
512$\times$512 computational nodes. 

We performed simulations for different
numbers of resonators per wavelength and different values of the ratio $\Omega = \omega_r/\omega$. The different cases 
are studied by changing the proper frequency and number of the oscillators, while keeping the amplitude and the 
length of the initial sinusoidal wave unchanged. This choice prevents a priori any change in the wave steepness which 
would in turn affect the nonlinearity of the wave dynamics.

\section{Results\label{sec:results}}

We first consider a single resonator placed at the centre of the domain where a monochromatic wave of wavelength $\lambda=1$ 
(in non dimensional units) propagates. In  figure \ref{fig:1pvis} we report the snapshot of the horizontal velocity component
$u$, the interface location and the oscillator position for the case $\Omega = 1$ at four instants of time,  $t/T = 0.5,1,1.5,2$, 
with $T$ the wave period. Wave motion forces the resonator to move due to pressure and viscous stresses distribution; the 
resonator, then, is pulled back to its original position by the elastic force and starts to oscillate around its 
equilibrium position. This motion induces perturbations on the interface, clearly visible at later stage of the process 
(figure \ref{fig:1pvis}d), leading to vorticity production at the surface which enhances the energy dissipation.

\begin{figure}
  \centering
  \includegraphics[width=0.8\columnwidth]{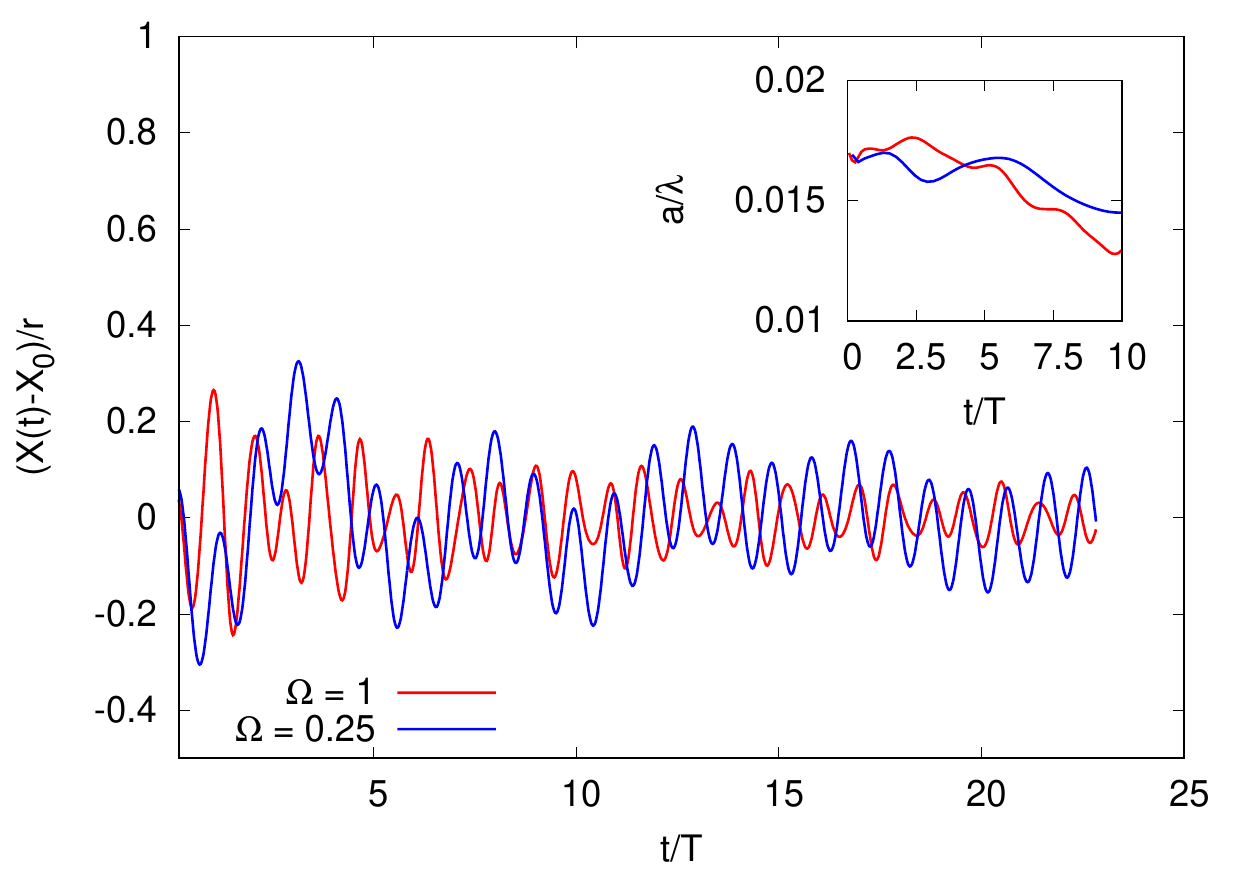}
  \caption{Time history of the center of mass displacement of one resonator placed in the middle 
    of the domain for two different values of $\Omega$; the inset shows the wave amplitude vs. time for the same cases.\label{fig:timeseries}}
\end{figure}

In figure \ref{fig:timeseries} we show the time history of the displacement of the centre of mass
of the resonator for $\Omega=1$, i.e. when the frequency of the wave is about the frequency 
of the resonator, and for $\Omega=0.25$, i.e. the frequency of the wave is 4 times the 
frequency of the resonator. In one period the wave has travelled the full domain.
Due to the periodic boundary conditions, the wave re-enters the domain from the left with a 
reduced amplitude, both because of its natural decay due to viscosity and because of the 
interaction with the resonators. Therefore, the resonator oscillates with a decreasing 
amplitude (as highlighted  in figure \ref{fig:timeseries}), since they
are forced by waves whose amplitude is decreasing in time. In the inset of the figure we plot 
the wave amplitude (computed as the difference between the maximum and minimum value of the surface elevation)
vs. time for the same cases: the simulation with $\Omega = 1$ exhibits a stronger 
decrease of wave amplitude which is in line with an increase of dissipation, as discussed below.

\begin{figure}[!ht]
  \centering
  \begin{subfigure}{\columnwidth}
    \includegraphics[width=\columnwidth]{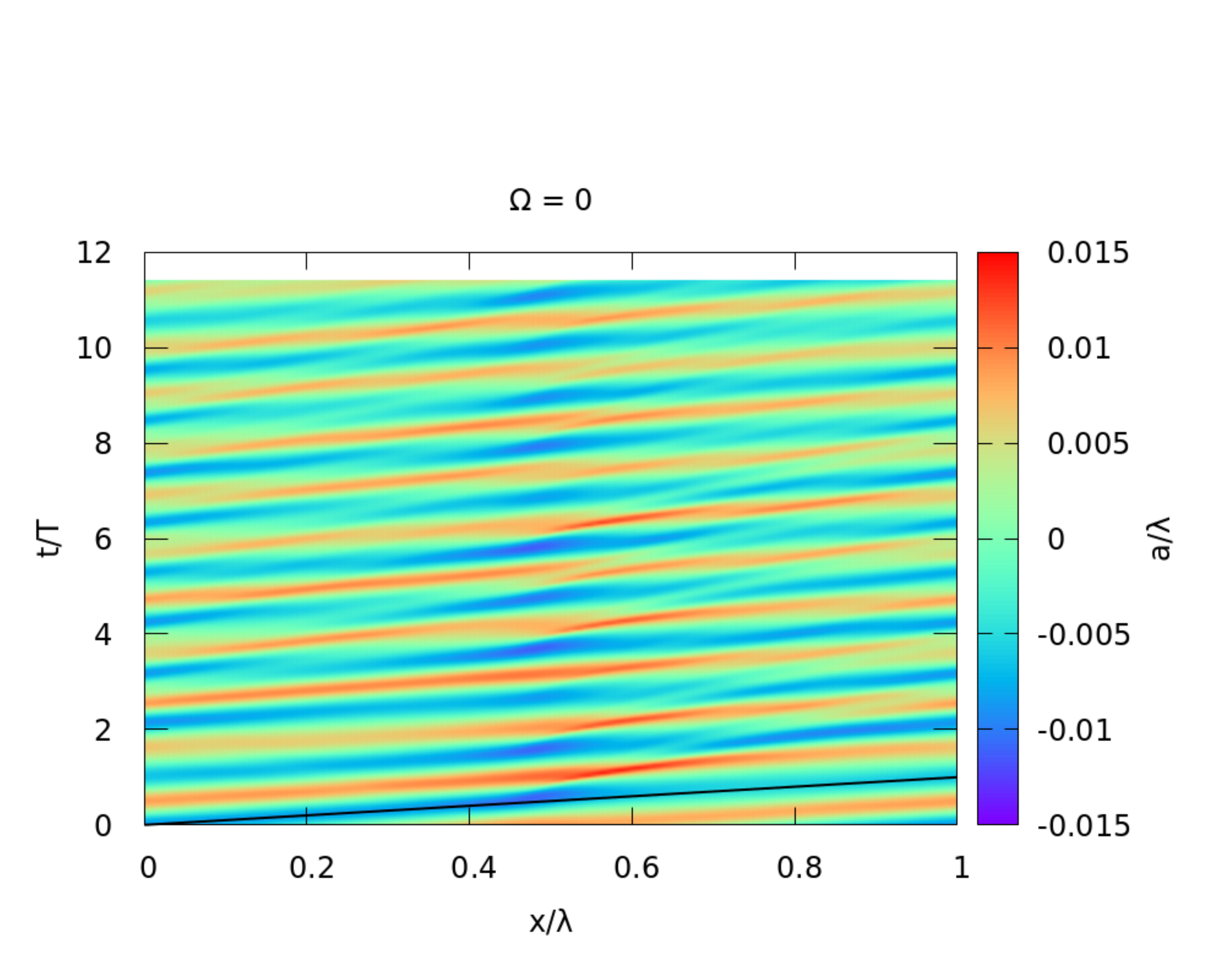}
  \end{subfigure}
  \begin{subfigure}{\columnwidth}
    \includegraphics[width=\columnwidth]{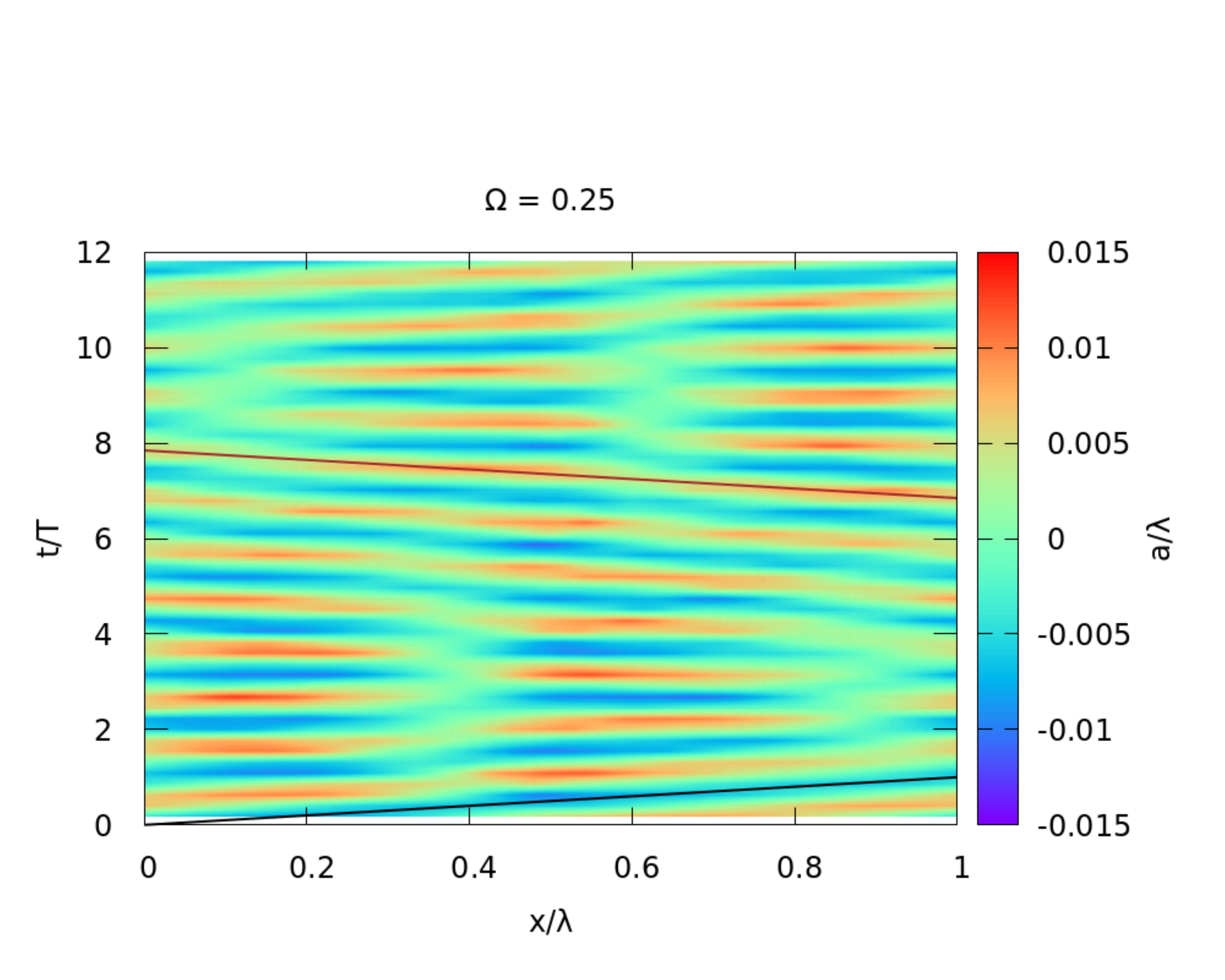}
  \end{subfigure}
  \begin{subfigure}{\columnwidth}
    \includegraphics[width=\columnwidth]{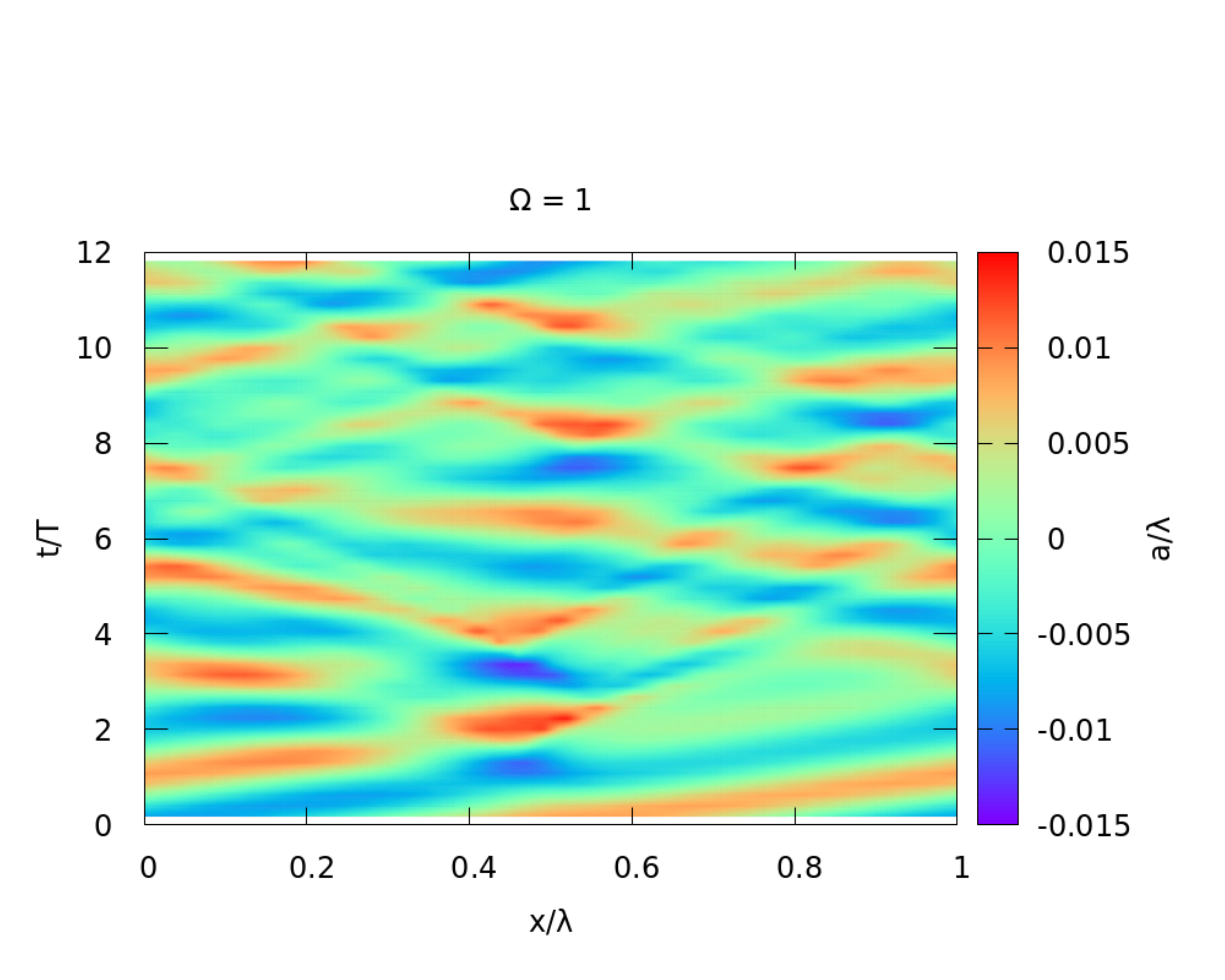}
  \end{subfigure}
  \caption{Space-time evolution of the surface elevation: (top) $\Omega = 0$;(middle) $\Omega = 0.25$; (bottom) $\Omega = 1$. The two  
  lines in the middle panel highlight forward (black line) and backward (brown line) propagating waves.\label{fig:conteta}}
\end{figure}

We find very instructive to show the space-time plots of the surface elevation, displayed
in figure \ref{fig:conteta}. The presence of the resonator induces a local perturbation of the surface elevation; 
this is clearly visible in the top panel of figure \ref{fig:conteta}, corresponding to the case with a 
fixed cylinder located at $x/\lambda = 0.5$. Additionally, when the solid body oscillates, the interaction with the 
propagating surface gravity wave leads to the generation of a wave travelling in the opposite direction with respect to the 
original one. This is highlighted in the middle panel of figure \ref{fig:conteta} by a brown line. For this simulation the period of the resonator is 
four times the wave period ($\Omega = 0.25$) and after approximately four non-dimensional times there is an inversion of the 
direction of propagation of the wave, which is again recovered after four more wave periods. For the case $\Omega = 1$, bottom panel of
figure \ref{fig:conteta}, a similar dynamics takes place on a shorter time scale but the evolution of the free-surface is less regular. 

In the following we will quantify the dissipated power during the wave propagation 
as a function of $\Omega$ and the number of resonators per wavelength. 
Due to the non-stationary nature of the system,  such measure must be described in a time-dependent fashion.
When a surface gravity wave of small amplitude (\emph{i.e.} small steepness 
$\varepsilon = a k$) propagates freely, its total energy decays with an exponential rate 
equal to $E(t) = E(0)e^{-2\gamma t}$, as described by \citeauthor{Landau}
\cite{Landau}. Here, the wave energy $E(t)$ is the sum of the
kinetic and potential contribution defined as:
\begin{equation}
  \begin{aligned}
    E(t)=&K(t)+U(t)=\\
    &\frac{1}{2}\int_0^\lambda\int_{-h}^{\eta}\rho{|\bf u|}^2 d z dx+
\int_{0}^{\lambda}\int_{-h}^{\eta}\rho g z dzdx - \bar{U}
  \end{aligned}
\end{equation}
where $z=-h$ is the position of the flat bottom, $z=\eta(x,t)$ is the displacement of the surface
with respect to its equilibrium position and $\bar{U} = \int_0^\lambda\int_{-h}^{0} \rho g z dzdx = -\rho g \lambda h^2$
the potential energy of the still water level.
 $E(0)$ is the initial energy budget of the wave and
$\gamma = 2\nu k^2$ is the decaying rate, $\nu$ being the kinematic viscosity of the fluid.
As mentioned, our aim is to evaluate the effect of the resonator on the propagation of the wave for
different values of the frequency of the resonator $\omega_r$. If the oscillator were in vacuum 
or in a low density fluid, its frequency would simply be given by $\omega_r=\sqrt{\kappa/m}$; however, because of the 
presence of a dense fluid, a proper  evaluation of  the latter 
needs to account also for the added mass given by the
surrounding fluid which results in a frequency $\omega_r = \sqrt{\kappa/(m + \rho_w\mathcal{V})}$,
with $\mathcal{V}$ the volume of the resonator. 

We start first by studying the interaction of
a single resonator with the wave. Depending on the ratio $\Omega$, the energy transfer
from the wave to the resonator can be more or less effective. 
\begin{figure}
    \centering
    \includegraphics[width=0.8\columnwidth]{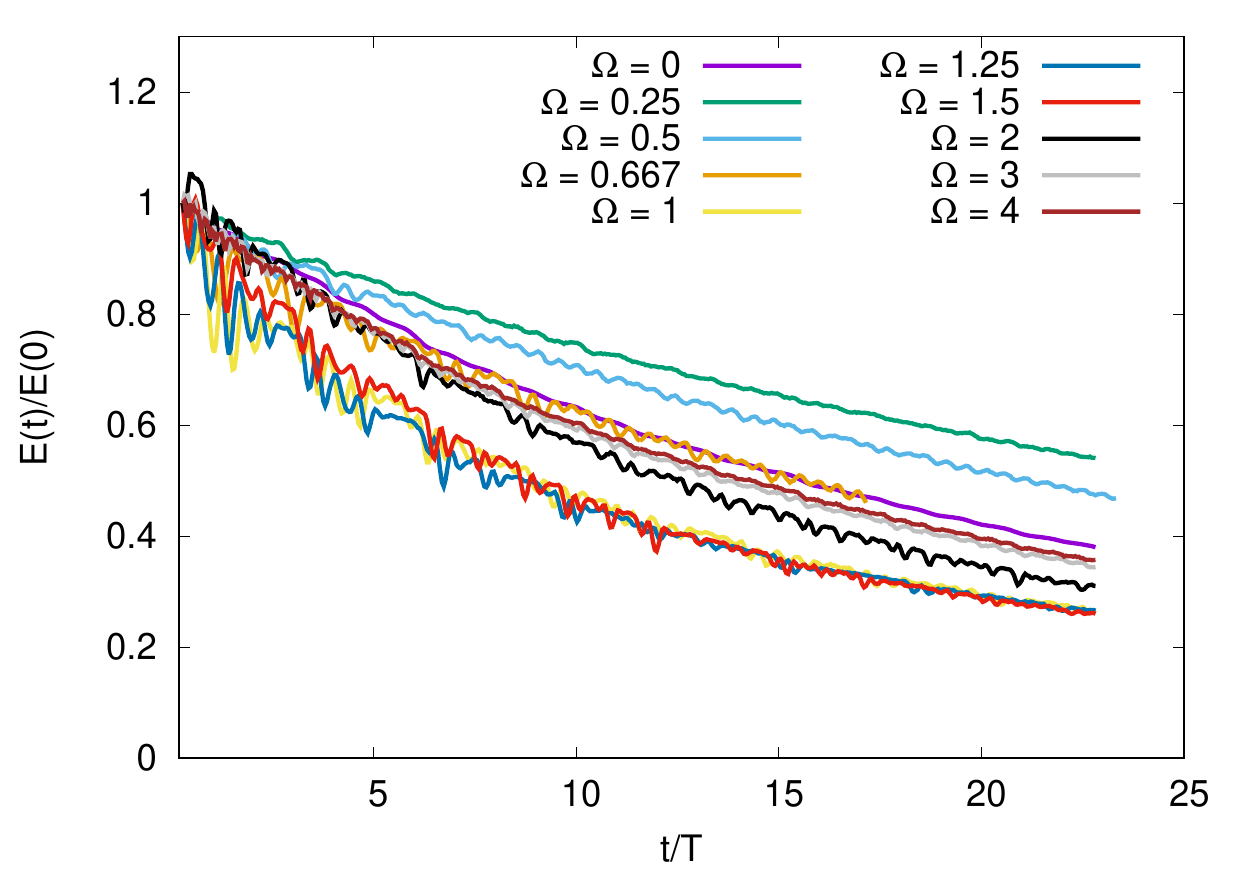}
    \caption{Time history of the total energy vs the frequency ratio $\Omega$ for the case of 
    one single resonator.\label{fig:moreom}}
\end{figure}
\begin{figure}
  \centering
    \includegraphics[width=0.8\columnwidth]{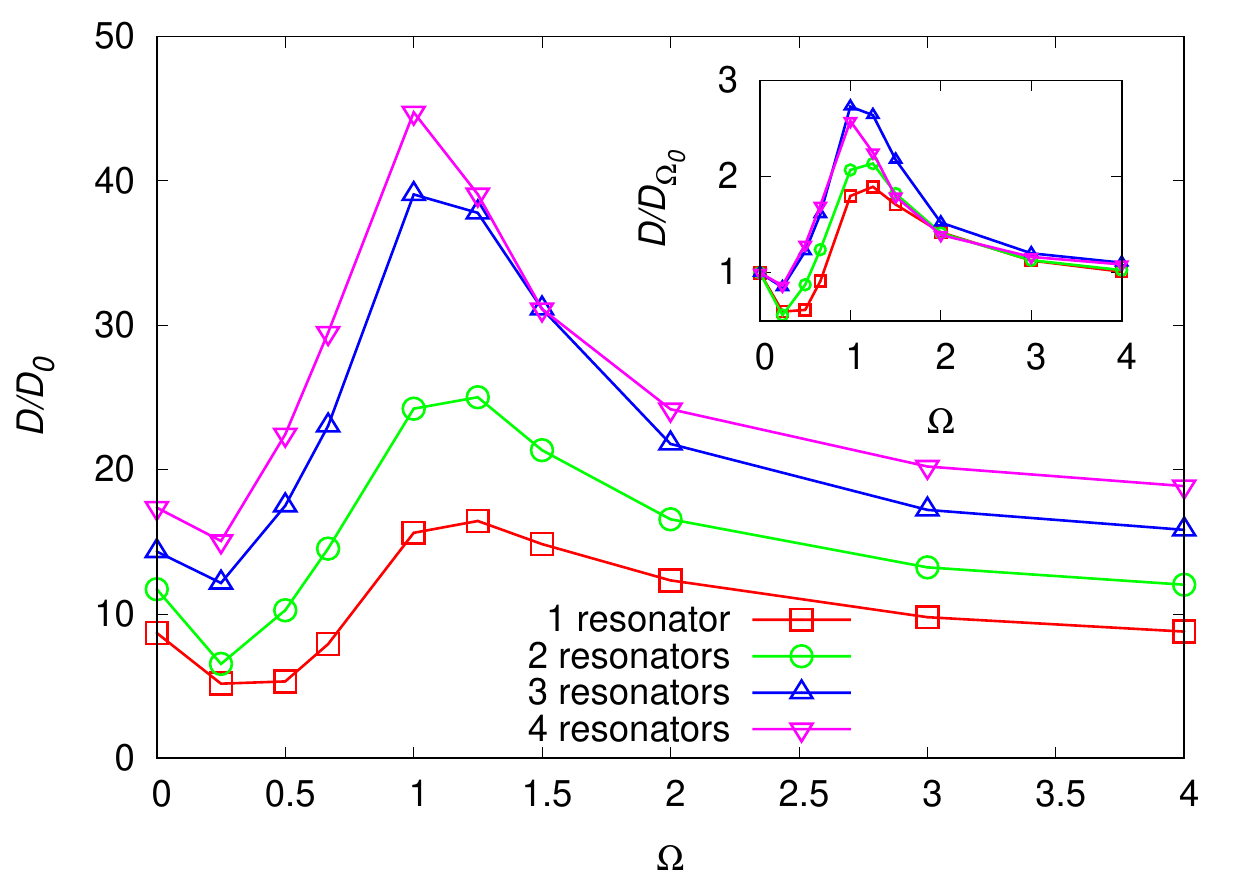}
    \caption{Dissipation coefficient $D$ normalised with the value for simple travelling wave
    $D_0$ vs the frequency ratio $\Omega$ for different number of oscillators. In the inset
    the same data are reported normalised with respect to the dissipation corresponding to the case of 
    fixed cylinders $D_{\Omega_0}$ (computed separately for each case).\label{fig:dissipation}}
\end{figure}
The effectiveness of the energy transfer from waves to the resonators is more clearly visible in the total 
energy history of the wave displayed in figure \ref{fig:moreom}.
The plot shows that small and large values of $\Omega$ are associated with
smaller dissipation, whereas values of $\Omega$ close to unity are more dissipative. In these
cases, a large amount of energy is dissipated in few wave periods. To quantify the 
dissipation we have performed a fit of these curves with an exponential form $E(t)\sim E(0)\exp[-Dt]$. 
The coefficients of the fit are reported in figure \ref{fig:dissipation}, where we show the results for simulations with up to 4 resonators.
The curves show  a clear peak of dissipation around $\Omega = 1$. 
The plots include the results for arrays of fixed cylinders, labelled by
$\Omega = 0$.
Independently of the number of resonators, we  observe a range of frequency ratios, between 0 and 1, for which
the dissipation is smaller than for the case of fixed resonators. Such dip is particularly evident for the case with one resonator per 
wavelength and is reduced when the number of cylinders is increased. The
dissipation then increases and has a peak for $\Omega\sim 1$. For $\Omega> 1$ it decreases and appears to approach an asymptote close
to the value found for fixed cylinders. It is worth noticing that the curves in 
figure \ref{fig:dissipation} resemble the curves reported for a system for wave energy conversion 
\cite{anbarsooz2014} in which the focus is on maximising the extracted power. This seems to indicate that the system of 
submerged resonators is a rather efficient system in dissipating wave power.
As the number of resonators is increased, the dissipation also increases. However, this is not a trivial effect due to the total viscous 
drag of the cylinders on the fluid. Indeed, the ratio between the peak value of the dissipation (around resonance) and the fixed-cylinder 
value increases as the number of resonator increases between 1 and 3, and appears to decrease with 4 resonators. This suggests that non-trivial 
interaction effects are present. It is also interesting to observe that 
the width of the dissipation peak, and therefore the range of frequencies
for which the dissipation is greater than the fixed-obstacle case, becomes
wider. 
\begin{figure}
  \centering
  \includegraphics[width=0.8\columnwidth]{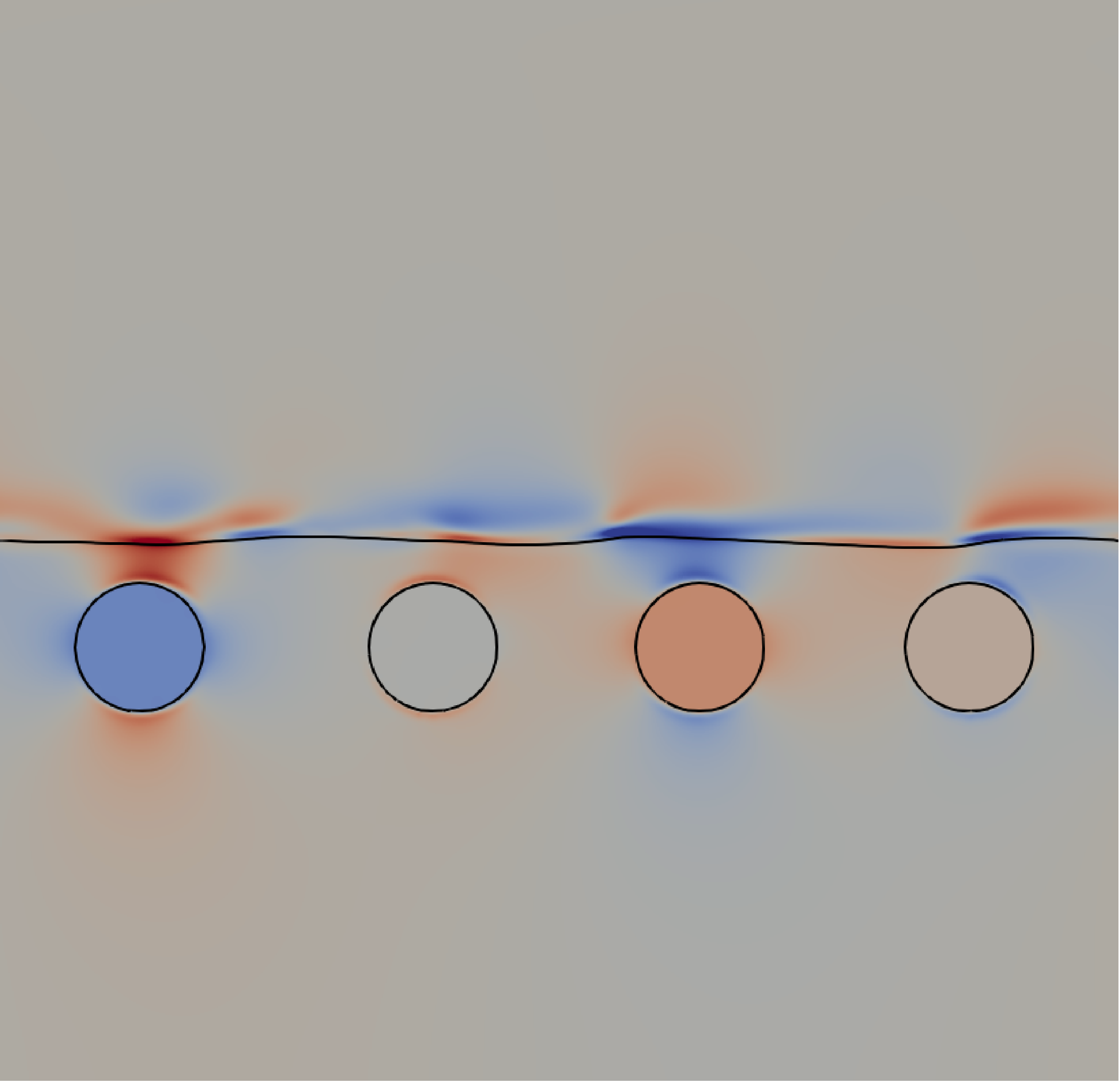}
  \caption{Snapshots of the horizontal velocity component $u$, interface location and resonators position
  for the case with four resonators and $\Omega = 1$ after one wave period. Colors as in figure \ref{fig:1pvis}. (Multimedia view)\label{fig:4pvis}}
\end{figure}
Therefore, decreasing the ratio between the wavelength of the wave and the wavelength of the periodic structures 
enlarges the range of frequencies for which an array of resonators produces a gain in dissipated power with respect to an array of fixed obstacles. 
An example of the flow field with four resonators and $\Omega = 1$ is reported in
figure \ref{fig:4pvis}. In this case the characteristic size of the perturbations induced by the resonators
on the interface is of order of the size of the periodic structure. 
It is worth mentioning that, in the presence of a current, two main effects could be expected:
\emph{i)} beacuse of the Droppler shift, the frequency of the wave can be shifted with respect to the one without 
a current; this would lead simply to an horizontal shift of figure \ref{fig:dissipation}; \emph{ii)} The current may result in an extra 
force on the cylinder due to the exchange of the momentum between the current and the resonator. Clearly if the current is small compared 
to the velocities induced by the waves, the effect is negligible. However, in the case of strong current, the cylinder may enter into an overdamped 
regime and may not oscillate anymore. However, the present model can handle the presence of a current, since it will be included in the rhs of 
\eqref{eqn:Newton}

\begin{figure*}
  \centering
  \begin{subfigure}{0.49\textwidth}
    \includegraphics[width=\textwidth]{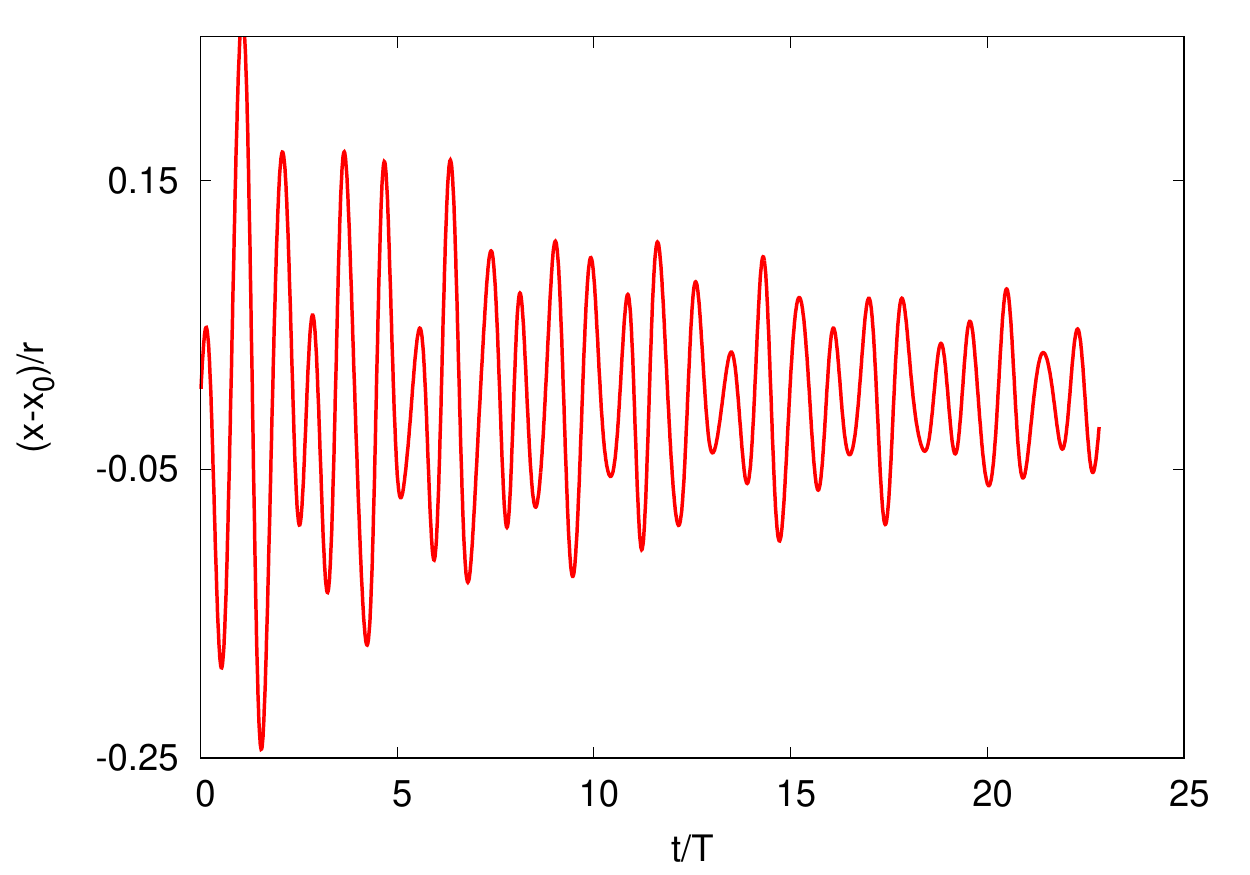}
    \caption{}
  \end{subfigure}
  \begin{subfigure}{0.49\textwidth}
    \includegraphics[width=\textwidth]{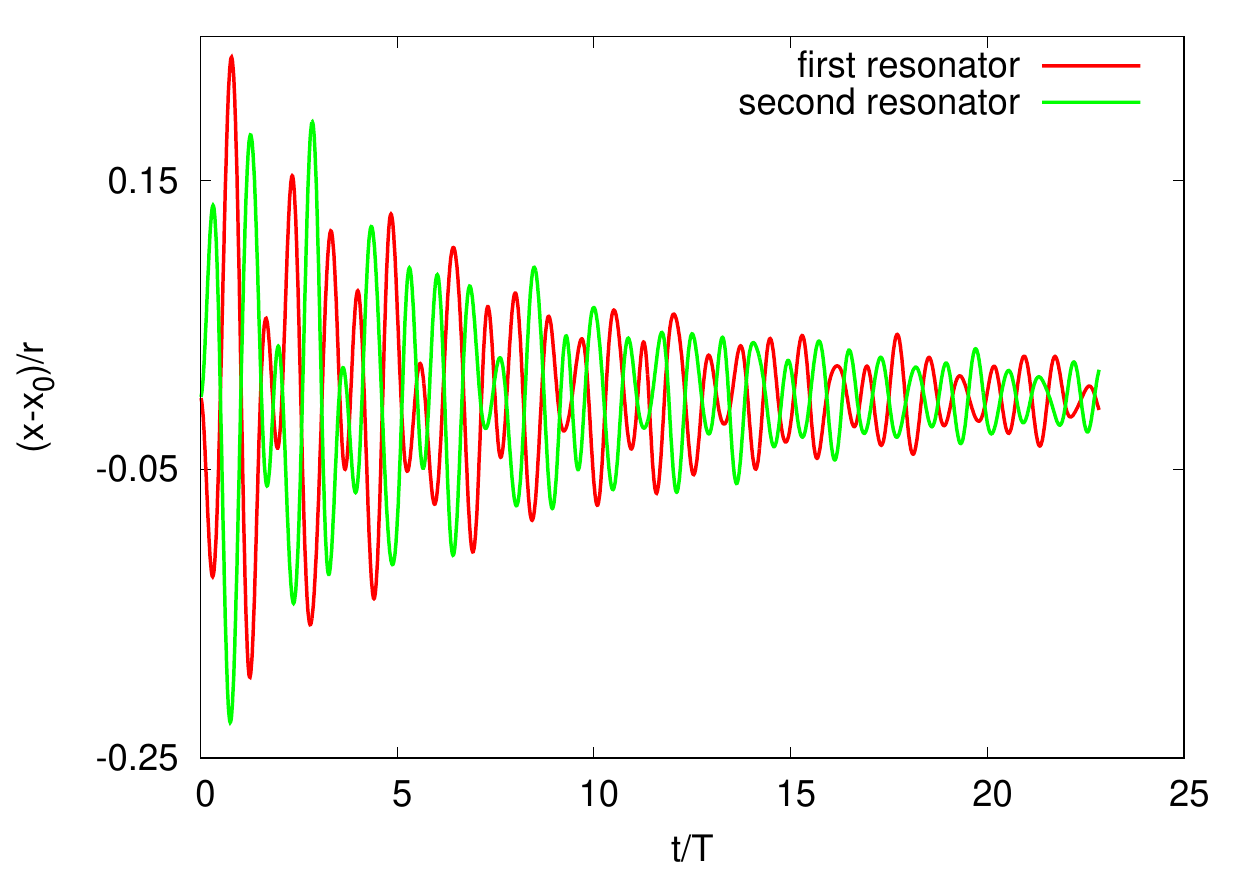}
    \caption{\label{fig:5b}}
  \end{subfigure}
  \begin{subfigure}{0.49\textwidth}
    \includegraphics[width=\textwidth]{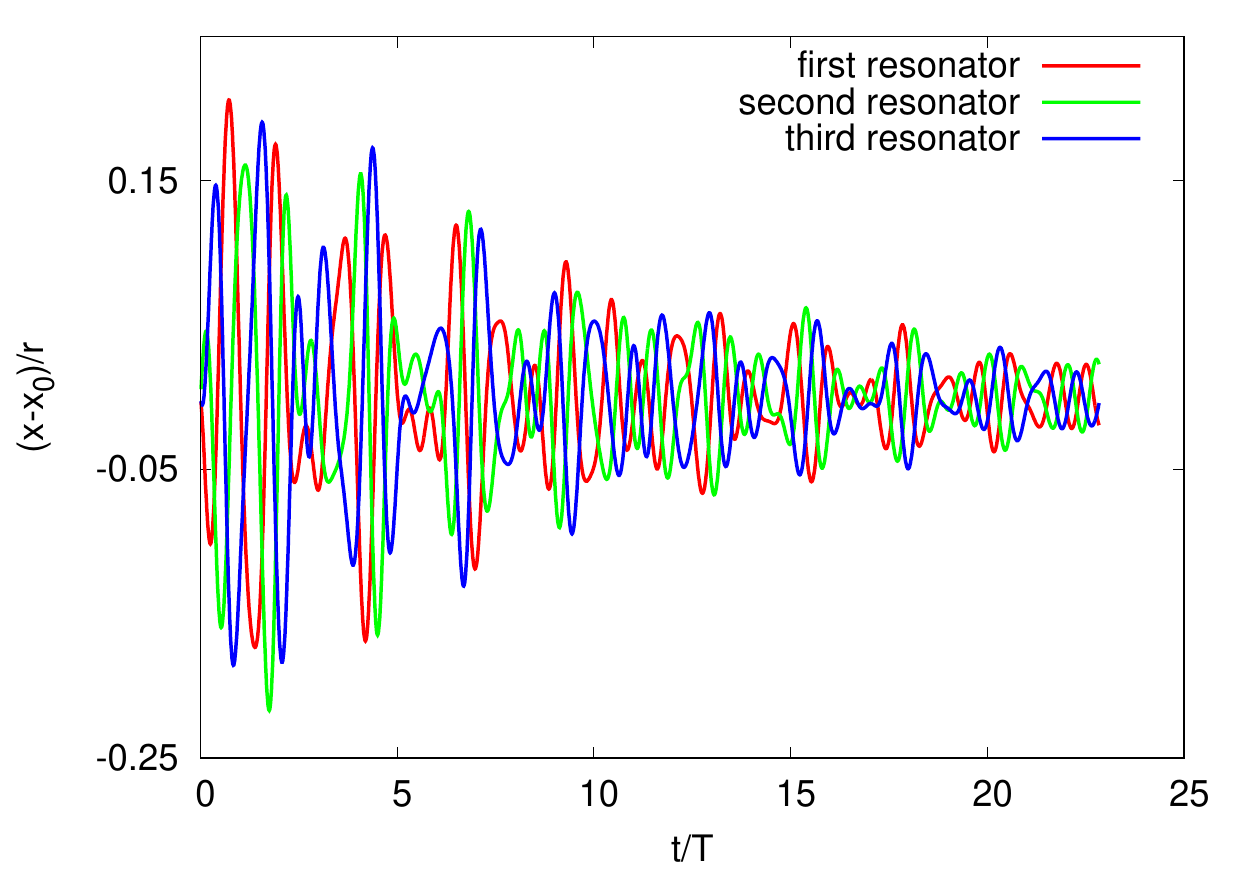}
    \caption{\label{fig:5c}}
  \end{subfigure}
  \begin{subfigure}{0.49\textwidth}
    \includegraphics[width=\textwidth]{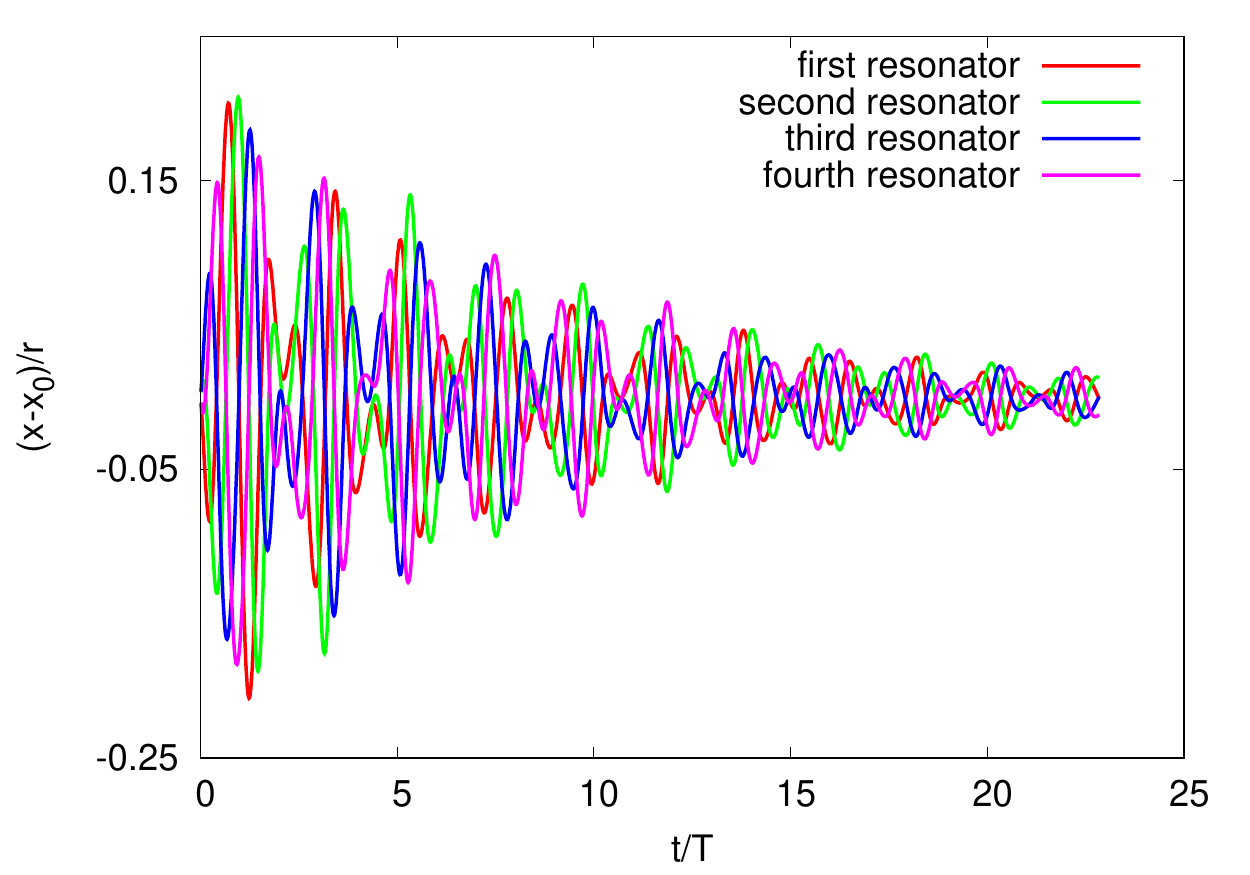}
    \caption{\label{fig:5d}}
  \end{subfigure}
  \caption{Time history of the horizontal position of the centre of mass of the
           oscillators: (a) case with 1 resonator (b) case with two resonators;
           (c) case with three resonators; (d) case with four resonators. For all cases
           $\Omega = 1$. \label{fig:correlation}}
\end{figure*}
In figure \ref{fig:correlation} we report the time history of the centre of mass of the resonators for 
$\Omega = 1$ and for different number of resonators. In the case of two oscillators (figure \ref{fig:5b}), the 
curves have a Pearson correlation index of about
-0.94 indicating that the oscillators are in phase opposition, as also clearly shown by the plot. For the case of 
three oscillators, instead, the correlation indexes of the curves are all about -0.5 because of a phase 
shift in the motion of the resonators (figure \ref{fig:5c}). Finally, in the last case we find that the oscillators
are correlated over a distance equal to $\lambda /2$, since the correlation coefficients for the first and third 
resonator, as well as that for the second and fourth, are about -0.97, while for the other pairs the coefficient is about 0.1.

\section{Conclusions\label{sec:discussion}}
In this work, we have proposed to exploit the concept of mechanical metamaterials 
in the field of fluid mechanics, using of submerged
resonators that interact with travelling waves, absorbing and dissipating
mechanical energy. In order to properly describe the behaviour of the system, we
have simulated the full Navier-Stokes equations for multiphase flows with
fluid-structure interaction; this approach allows for a complete and detailed
evaluation of hydrodynamic forces acting on the resonators and of the energy
dissipation.

We have performed simulations in a periodic square domain of
size equal to the wavelength of the wave, varying the elastic force acting on
the resonators (\emph{i.e.} their natural frequency) and the number of resonators
per wavelength. We have computed the time
history of the wave energy and found a dissipation coefficient by fitting the
energy decay with an exponential form, similar to that of the viscous
dissipation of a simple travelling wave. By doing so, we have found that there
is a peak of dissipation when the frequency of the wave and the frequency of
the resonators approximately coincide. The dissipation observed at the peak is much
larger that that caused by an array of fixed cylinders, so that the width of
the peak represents the range for which the oscillatory dynamics (and, possibly, the fluid-mediated 
interaction among the structures) produces a gain in the dissipated power.
Finally, the presence of a dissipation peak centred around a characteristic frequency suggests the 
presence of a band gap in the dispersion relations. The width of such band gap should increase with the number of resonators.

Future work will focus on coupling this system with a numerical wave maker to properly evaluate the dispersion 
relation and also to investigate the effect of resonator masses on the band gap. Additionally, 
the extension of the method to deformable solid bodies could open the field of applications also to flexible 
underwater structures.
This work could open 
new applicative possibilities to realize low-cost, minimally invasive devices for ocean wave attenuation, 
contributing to reduced costal erosion or protection of infrastructure such as offshore platforms or harbours.

\section*{Data Availability}
The data that support the findings of this study are available from the corresponding author upon reasonable request.

\section*{ACKNOWLEDGMENTS}
M. Onorato, F. De Lillo and F. De Vita have been funded by Progetto di Ricerca
d'Ateneo CSTO160004, by the ``Departments of Excellence 2018/2022''
Grant awarded by the Italian Ministry of Education, University and
Research (MIUR) (L.232/2016).
The authors acknowledge the EU,
H2020 FET Open BOHEME grant No. 863179 and 
CINECA for the computational resources under the
grant IscraC SGWA.

\nocite{*}
\bibliography{biblio}

\end{document}